\documentclass[12pt,preprint]{aastex}

\begin{document}

\title{Dusty Cometary Globules in W5}

\author{X.~P. Koenig,\altaffilmark{1} L.~E. Allen,\altaffilmark{1} S.~J. Kenyon,\altaffilmark{1} K.~Y.~L. Su,\altaffilmark{2} Z. Balog\altaffilmark{2}}
\altaffiltext{1}{Harvard-Smithsonian Center for Astrophysics, 60 Garden Street, 
Cambridge, MA 02138}
\altaffiltext{2}{Steward Observatory, University of Arizona, 933 North Cherry Avenue, Tucson, AZ 85721}

\begin{abstract}
  We report the discovery of four dusty cometary tails around low mass
  stars in two young clusters belonging to the W5 star forming
  region. Fits to the observed emission profiles from 24 $\micron$
  observations with the {\it Spitzer} Space Telescope give tail
  lifetimes $<$ 30 Myr, but more likely $\lesssim$ 5 Myr. This result
  suggests that the cometary phase is a short lived phenomenon,
  occurring after photoevaporation by a nearby O star has removed gas
  from the outer disk of a young low mass star \citep[see
  also][]{balog06,balog08}.
\end{abstract}


\section{Introduction}
Recent observations with the Multiband Imaging Photometer for {\it
  Spitzer} (MIPS) have discovered dusty nebulae associated with
several young stars in three Galactic star forming regions
\citep{balog06}. Appearing as extended, cometary objects pointing away
from nearby O stars at 24 $\micron$, (and sometimes at 8 $\micron$),
they resemble the proplyds seen in Orion with HST
\citep{odell93,odell94}. \citet{balog06} show that these cometary
tails are probably dust swept out of a young circumstellar disk by an
O star.

\citet{balog06} show that the emission profiles at 24 and 8 $\micron$
can be explained by a model where gas is photoevaporated from a young
protoplanetary disk. As the gas leaves, entrained dust also escapes
and is then blown away from the disk by radiation pressure from nearby
O stars. The dust is swept into a cometary morphology which glows in
the mid-IR thermally as it reprocesses the incident UV radiation to
longer wavelengths. \citet{balog06} cite \citet{johnstone98},
\citet{richling00}, \citet{hollenbach04} and \citet{matsuyama03}, who
find that the portion of a protoplanetary disk likely responsible for
emission at 24 $\micron$ ($\sim$2--50 AU from the star) is removed on
timescales $\sim$3$\times 10^5$ yr.

We have recently discovered four dusty cometary structures in the W5
star forming region. Here we expand on the work of \citet{balog06} and
place constraints on the lifetimes and physical nature of these
interesting objects.

\section{Observations}
\citet{koenig08} describe our observations of W5. We observed a
2$\degr \times$1.5$\degr$ field with the Infrared Array Camera (IRAC)
and MIPS in 2006 and 2007. Figure 1 shows the MIPS 24 $\micron$ image
mosaic. Images were taken in scan map mode using the medium scan speed
for an average exposure time of 41.9 seconds per pixel for the
combined frames. The raw MIPS data were processed with pipeline
version S13.2.0. The final mosaics were produced using the MIPS
instrument team Data Analysis Tool, which calibrates the data and
applies a distortion correction to each individual exposure before
combining \citep{gordon05}. We used the {\it clustergrinder} software
tools \citep[described in][]{gutermu08} to produce final image mosaics
from the IRAC data in each of the four wavelength bands.

Figure 2 shows the 24 $\micron$ cometary objects at larger scale. At
wavelengths shorter than 24 $\micron$, these objects are point-like or
not detected. At 70 and 160 $\micron$, these objects are also
invisible. Table 1 summarises the properties of the four objects in
W5. The quoted 24 $\micron$ flux is for the heads of the cometary
features alone.

\section{Analysis}
We follow \citet{balog06} and analyze these objects using an
adaptation of the Vega debris disk model of \citet{su05}. We assume
the outward flow has roughly constant radial density: $n(r) \sim$
constant. We fit the emission profile of a 5 pixel (6.23$\arcsec$)
wide strip along the symmetric axis of the tail. To estimate and
subtract the background, we extend the strip to reach regions free of
diffuse emission and point sources.

The fit to the resultant surface brightness profile is based on
optically thin emission from dust at its equilibrium temperature. The
emission depends on dust composition, dust grain size, and the angle
that the tail makes with the plane of the sky (inclination), where
0$\degr$ corresponds to a tail in the plane of the sky. We adopt
astronomical silicates for the dust composition. Due to the degeneracy
between tail inclination and dust grain size \citep{balog06}, and the
lack of reliable constraints on these parameters, we assume an
inclination of 0$\degr$ and fit the profile adopting a single grain
size.

The fits to the 24 $\micron$ emission profiles are shown in Figure
3. Error bars on the profile points are derived by extracting the
emission profile to either side of the main axis and taking the
difference between the two extractions as the error at each
point. Tails A, B and C are formed by the multiple O star system HD
17505 which contains at least four O stars \citep{hillwig06}. We model
the radiation output of this system as a single star with effective
temperature $T_{eff}$ = 35000 K and radius $R = 21 R_{\odot}$. Tail D
is most likely formed by the O7V star HD 18326. We adopt $T_{eff}$ =
36000 K and $R = 9.4 R_{\odot}$. The model provides reasonable fits to
the dust emission profiles at 24 $\micron$, however all of the fits
diverge from the data at either end due to contamination from nearby
stars.

At 8 $\micron$, our model predicts a flux of 0.0025 mJy for tail
A. This flux is consistent with the upper limit listed in Table 1. For
tails B and D the model underpredicts the 8 $\micron$ flux by more
than an order of magnitude. Tail C is a more complex case. Due to
crowding, we could not derive a precise flux for this object. The
model predicts a flux $>$5 times the upper limit quoted in Table 1 for
tail C. For tails B and D, the enhanced level of 8 $\micron$ emission
in our data is probably produced by stochastically heated small
grains, $\sim$10$\textrm{\AA}$ in size as suggested by
\citet{balog06}. To test whether this is also true for tails A and C
requires improved measurements of the flux at 8 $\micron$.

Optical and near infrared observations of the dusty tails in NGC 2264,
IC 1396 and NGC 2244 (Balog et al. 2008) suggest that these tail
structures are strongly deficient in gas. If gas does not entrain the
dust away from the circumstellar disk, another mechanism must produce
these structures. We propose that radiation pressure from nearby O
stars ejects the dust directly. To test this idea, we calculate the
radius, $r_{evac}$ within the stellar disk where radiation pressure
exceeds the gravitational binding energy or dust in orbit around its
host star \citep[see][]{burns79}. For a star of mass $m_{star}$:

\begin{equation}
  r_{evac} = \sqrt{\frac{8 G \rho \pi d^2\textrm{ } c \textrm{ } a\textrm{ } m_{star}}{3 L_O \overline{Q_{pr}}}} 
\end{equation}

\noindent where $d$ is the distance from the tail head to an O star
with luminosity $L_O$, and $\overline{Q_{pr}}$ is the frequency
averaged radiation pressure coefficient ($\overline{Q_{pr}}\approx$1
for an O star and astronomical silicate dust grains; Laor \& Draine
1993)\footnote{The expression for $r_{evac}$ assumes grains produced
  from collisions of larger objects. For primordial grains orbiting
  the central star, $r_{evac}$ is a factor of $\sqrt{2}$ larger.}. For
a 1 M$_{\sun}$ host star and grains of size $a$ = 0.01 $\micron$, and
density $\rho$ = 3.3 g cm$^{-3}$, $r_{evac}$ = 9--35 AU. For 10
$\micron$ grains, $r_{evac} \gtrsim$ 500 AU, the typical protostellar
disk radius \citep{andrews07}. Thus radiation pressure can eject
grains with $a \lesssim$3 $\micron$.

Table 2 lists two values for the tail mass. The lower limit assumes a
single grain size of 0.01 $\micron$. The upper limit assumes a size
distribution from 0.01 $\micron$ up to the grain size at which
$r_{evac}\sim350$ AU. We adopt the size distribution of
\citet{dohnan69} and \citet{tanaka96}, $N(a) \propto a^{-3.5}$ for a
grain population formed by a collisional fragmentation cascade in a
protoplanetary disk, normalize the mass in 0.1 $\micron$ grains
required to fit the surface brightness distribution (Fig. 3) and
integrate over grain size. Recently, \citet{jorda07} and
\citet{holsapple07} found a different power law ($-3.1$) for dust
grains in the Deep Impact ejecta experiment, which would result in a
factor $\sim$10 greater tail mass.

To derive a simple mass-loss rate for each tail, we divide the
estimated tail mass by the projected tail length over the flow
velocity \citep[following Balog et~al. 2008 we use the drift velocity
given by eqn. B.4 of][]{aberg02}. To derive an expected lifetime, we
adopt a minimum mass solar nebula as specified by \citet{weiden77} and
\citet{hayashi85} as a dust reservoir and find a total mass $M_{tot}$
= 0.034 M$_{\sun}$ between radii 20--350 AU. Assuming the dust mass is
a factor 100 smaller, the mass loss rates yield lifetimes of 0.3--30
Myr (Table 2). We note that \citet{andrews07} find a range in disk
masses around young stars, such that $M_{tot}$ could be as much as a
factor 10 larger.

\section{Discussion}
The initial phase of a young stellar disk's evolution is the
optically-thick phase. This period lasts a few to $\sim$10 Myr; during
this time, excess emission from dust in the circumstellar disk in the
{\it Spitzer} IRAC bands (3--8 $\micron$) is visible, as are
signatures of accretion onto the star \citep{aguilar05,aguilar06}. The
cometary tails in W5 are unlikely to be older than 5 Myr. Since
massive O stars are present and non-thermal radiation is largely
absent (which would be the signature of a supernova having previously
occurred), 5 Myr is a conservative age upper limit for W5
\citep{karr03,vallee79}. During this early phase, UV radiation from
nearby O stars will photoevaporate the outer disk of stars in their
clusters, creating proplyds such as those seen in Orion
\citep{odell93,odell94}. \citet{balog08} suggest that after this part
of the disk is removed, micron-size dust grains are replenished by
collisions between already-formed planetesimals and then blown out by
radiation pressure, creating the cometary tails seen at 24
$\micron$. We consider this scenario somewhat unlikely in such young
disks. \citet{kenyon01} find that a size distribution of bodies can
stir up planetesimals enough to begin a collisional cascade capable of
making small dust grains in the outer 30--150 AU of a disk, only if it
contains objects of 500 km radius or larger, which need at least
$\sim$10 Myr to form at this distance from the star
\citep{kenyon08}. Since we find that the dust grains that create the
tails come from beyond $\sim$10 AU they must have remained in the disk
and survived being completely dragged out as the gas departs during
photoevaporation.

We note that some of the objects could be older than the O stars and
have migrated in to their present location. This scenario requires an
earlier generation of stars (5--10 Myr ago) to have formed in a
cluster of $\sim$1000 stars within $\approx$5 pc of HD 17505 or HD
18326. In addition, faster planet formation \citep[e.g. by disk
instability, or induced by the radiative
environment][]{boss06,throop05} might allow the planetesimal stirring
and collisional cascade mechanism to operate and create the small dust
grains that make up the cometary tails. The dust may not remain long
enough in such a disk to make detection likely however
\citep{kenyon08}.

Tails A, B and C belong to the large double cluster of young stellar
objects (YSO) surrounding O stars HD 17505 and HD 17520. This cluster
has a projected radius on the sky of 3 pc, within which are 146 stars
classified by \citet{koenig08} as class II stars with disks,
i.e. stars in the optically thick disk phase, including tails A, B and
C. If the cluster is spherical and uniformly distributed we would
expect 1.54 objects within a spherical radius 0.67 pc (the distance of
tail A from HD 17505). Similarly tail D belongs to the cluster around
HD 18326 which has a radius of 1.8 pc and 87 YSO/optically thick disk
objects. Within a 0.22 pc radius (that would encompass tail D) we
would predict 0.16 objects. Given that projection effects mean we
should see only one third of these objects as extended, it is likely
that all objects with disks capable of producing cometary tails within
0.7 pc of an O star are doing so.

Not all O stars appear to create tails however. Within W5, 2 O star
systems (several are binary or multiple) out of 5 possess cometary
objects at 24 $\micron$. {\it Spitzer} observations of star forming
regions with MIPS (GTO program ID 58) have found 3 more out of a
sample of 20 O stars \citep{balog06}. This sample includes the Orion
Trapezium region which is saturated at 24 $\micron$ but does contain
the well known optically visible proplyds \citep{odell93}. The
timescale of the mid-infrared cometary phase may thus be closer to the
short end of our estimated range.

\section{Conclusions}
We have discovered four dusty cometary tails in W5 at 24 $\micron$
with {\it Spitzer} MIPS. An apparent deficit of gas suggests an origin
in radiation pressure blowout of dust from a young stellar disk by
nearby O stars. Future observations with high sensitivity, high
spatial resolution mid-infrared and sub-millimeter instruments will
help constrain their nature.

{\it Facilities:} \facility{2MASS ($JHK_S$)}, \facility{Spitzer (IRAC, MIPS)}


\clearpage

\begin{deluxetable}{lcccccccccc}
\tablenum{1}
\tablewidth{0pt} 
\tabletypesize{\tiny}
\tablehead{ \colhead{ } & \colhead{R.A.} & \colhead{Declination} & \colhead{$J$} & \colhead{$H$} & \colhead{$K_S$} & \colhead{$[3.6]$} & \colhead{$[4.5]$} & \colhead{$[5.8]$} & \colhead{$[8.0]$} & \colhead{$[24]$} \\ \colhead{ } & \colhead{J2000.0} & \colhead{J2000.0} & \colhead{(mJy)} & \colhead{(mJy)} & \colhead{(mJy)} & \colhead{(mJy)} & \colhead{(mJy)} & \colhead{(mJy)} & \colhead{(mJy)} & \colhead{(mJy)}}
\startdata 
A & 2 51 15.17 & 60 24 20.31 & \nodata & \nodata & \nodata & $<$0.006 & $<$0.007 & $<$0.035 & $<$0.048 & 19$\pm$3\\
B & 2 51 13.53 & 60 25 02.36 & \nodata & \nodata & \nodata & 0.40$\pm$0.04 & 0.27$\pm$0.02 & 0.15$\pm$0.05 & 0.19$\pm$0.05 & 16$\pm$2\\
C & 2 51 10.49 & 60 25 07.15 & \nodata & \nodata & \nodata & $<$0.006 & $<$0.007 & $<$0.035 & $<$0.048 & 26$\pm$2\\
D & 2 59 20.91 & 60 34 14.82 & 0.43$\pm$0.09 & 0.54$\pm$0.12 & 0.69$\pm$0.12 & 0.42$\pm$0.04 & 0.32$\pm$0.04 & 0.18$\pm$0.08 & 0.59$\pm$0.06 & 37$\pm$3
\enddata
\tablecomments{$JHK_S$ photometry obtained from 2MASS catalog \citep{skrut06}. We obtained the remaining photometry in {\it Spitzer} bands at tail head position, using aperture photometry.}
\end{deluxetable}

\begin{deluxetable}{lccccccc}
\tablenum{2}
\tablewidth{0pt} 
\tablehead{ \colhead{ } & \colhead{Tail Length} & \colhead{$r_{evac}$ (0.01$\micron$ grains)} & \colhead{$a_{max}^{\dagger}$} & \colhead{Tail Dust Mass$^{\ddagger}$} & \colhead{$v_{drift}$} & \colhead{Lifetime} \\
  \colhead{ } & \colhead{(pc)} & \colhead{(AU)} & \colhead{$\micron$} & \colhead{(10$^{-6}$
    M$_{\odot}$)} & \colhead{km s$^{-1}$} & \colhead{(Myr)}} \startdata
A & 0.3 & 34.5 & 1 & 0.15 -- 1.5 & 40 & 1.6 -- 15.8 \\
B & 0.49 & 20.2 & 3 & 0.08 -- 1.3 & 70 & 1.6 -- 30.1 \\
C & 0.62 & 9.3 & 10 & 0.06 -- 4.7 & 150 & 0.3 -- 24.5 \\
D & 0.29 & 23.6 & 3 & 0.14 -- 2.4 & 55 & 0.7 -- 12.6
\enddata
\tablecomments{$^{\dagger}$Maximum grain size evacuated from disk under assumption that no significant material exists beyond 350 AU. $^{\ddagger}$Tail dust masses for assumed single grain size and for power-law grain size distribution.}
\end{deluxetable}

\clearpage

\begin{figure}
\plotone{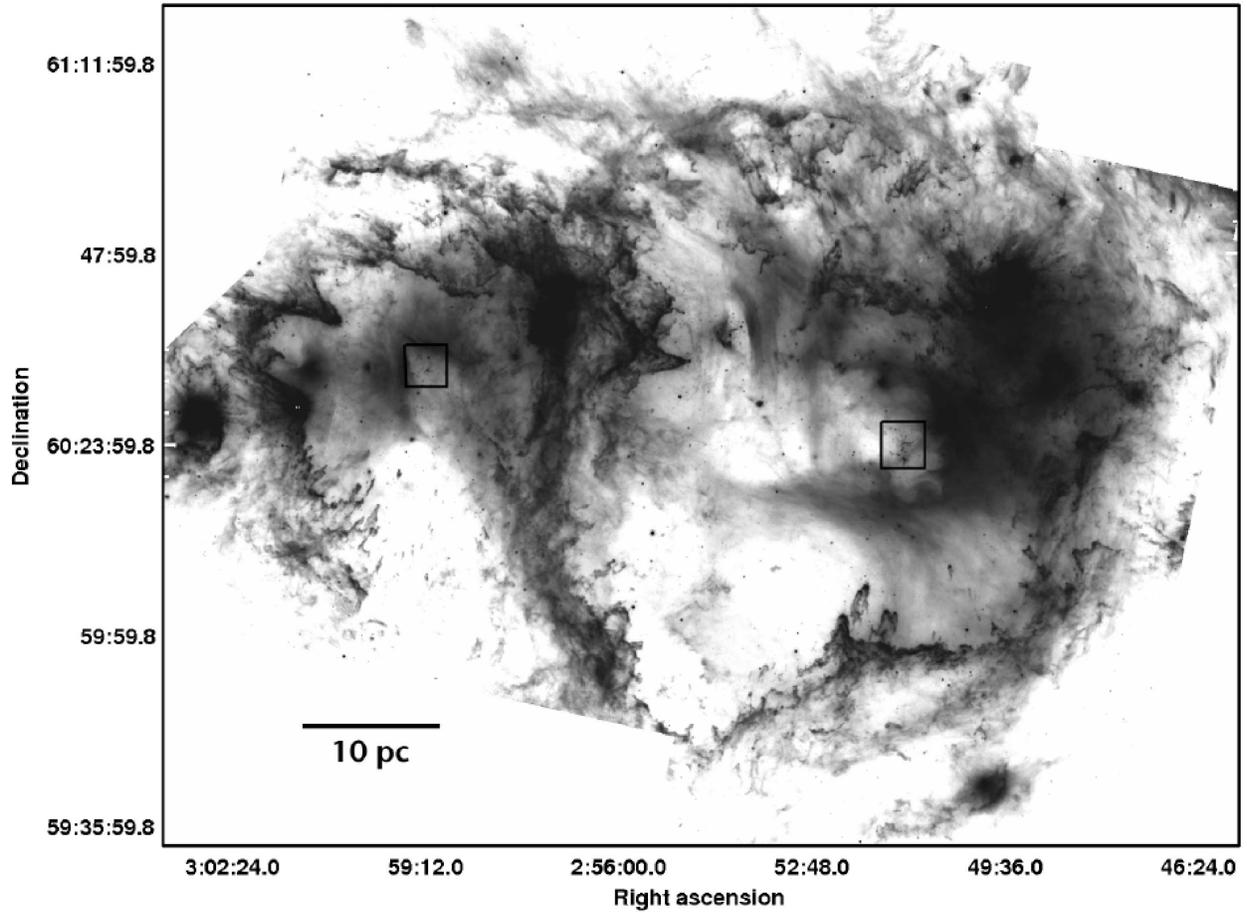}  
\caption{24 $\micron$ mosaic of W5, as imaged by {\it
    Spitzer}/MIPS. Two bubbles are seen in projection on the
  sky. Cometary sources are seen in the vicinity of the primary
  ionizing O stars. Expanded views of the 2 boxed regions are shown in
  Figure 2. The left hand box covers HD 18326, and the right hand box
  HD 17505.}
\end{figure}

\begin{figure}
\begin{center}
\includegraphics*[width=3.3in]{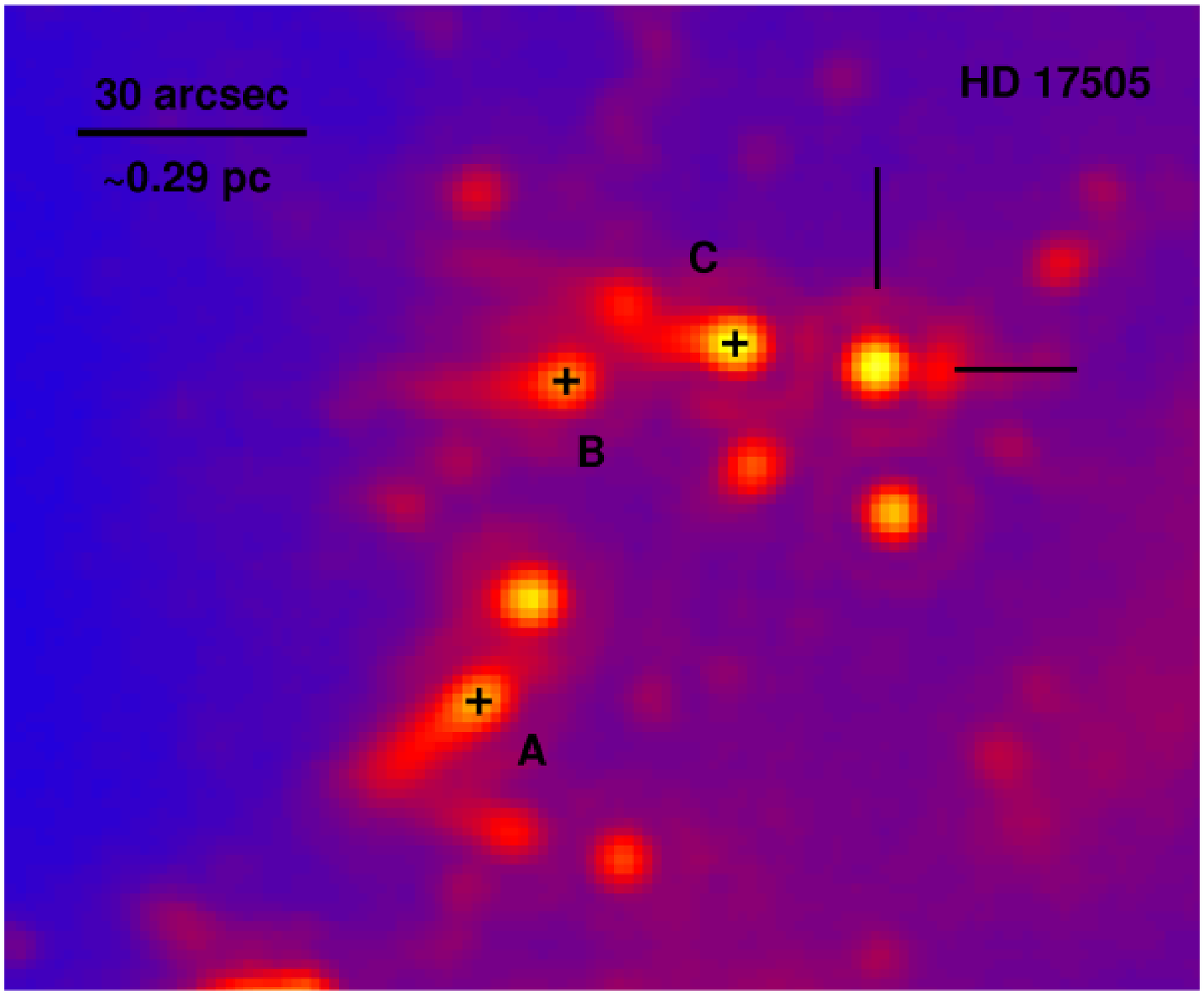}
\includegraphics*[width=3.3in]{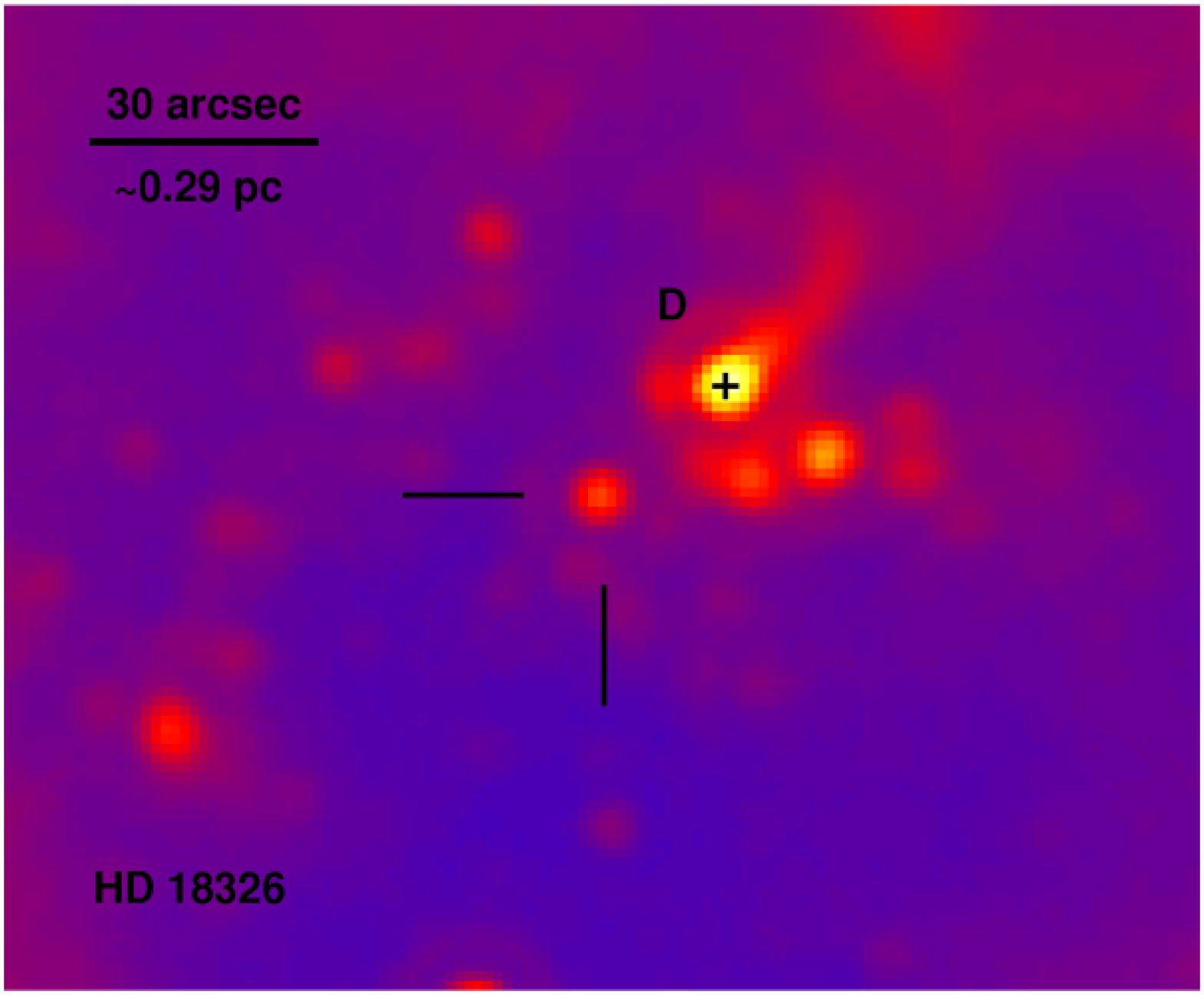}
\caption{Expanded views of the boxed regions in Figure 1 in
  exaggerated color to show the cometary sources (marked with crosses)
  and ionizing stars (marked by tics). Several cometary sources point
  to the multiple O star system HD 17505 (upper panel, see Hillwig et
  al. 2006). One cometary object is evident near HD 18326 (lower
  panel).}
\end{center}
\end{figure}

\begin{figure}
\begin{center}
\includegraphics*[width=3.2in]{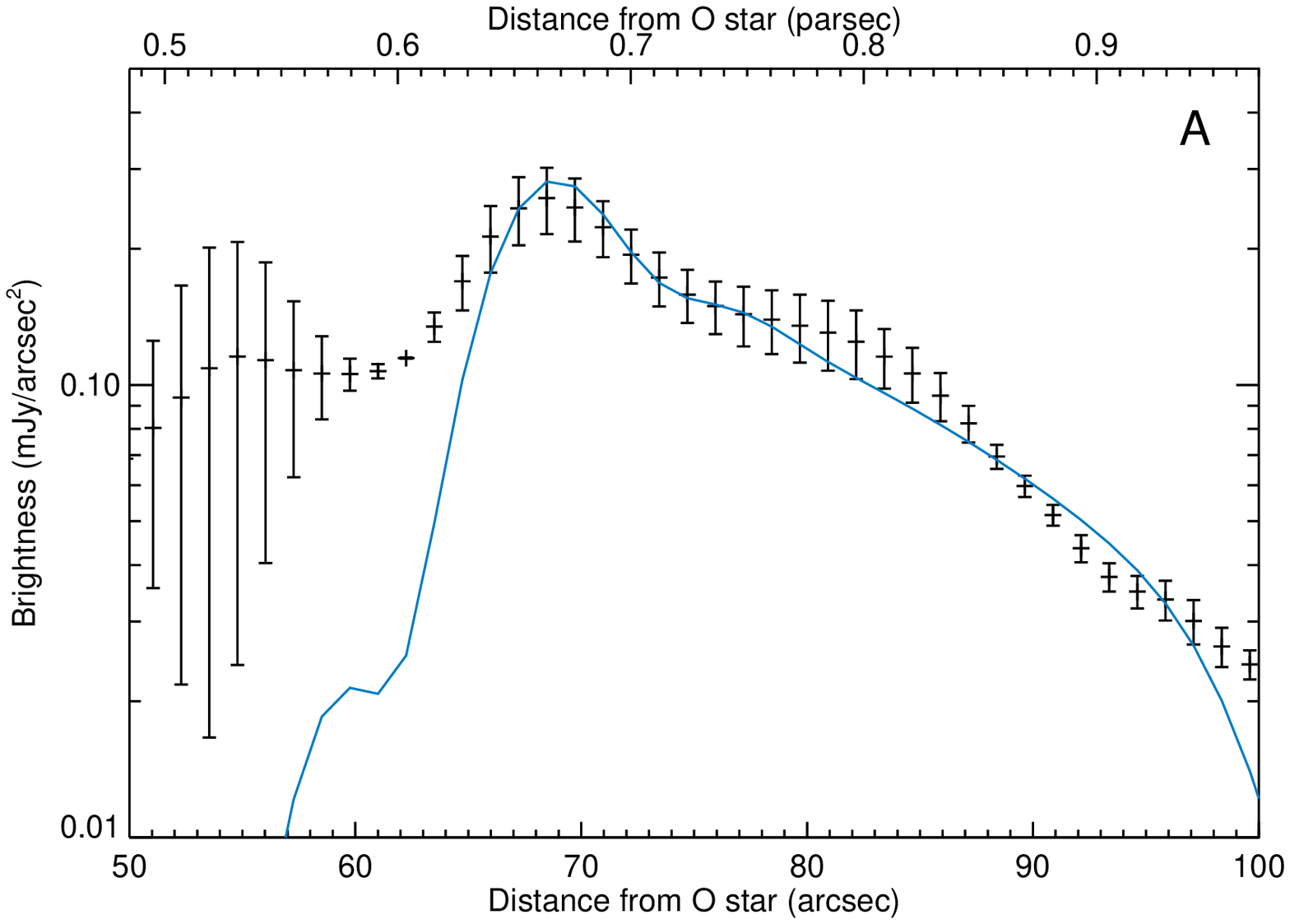}
\includegraphics*[width=3.2in]{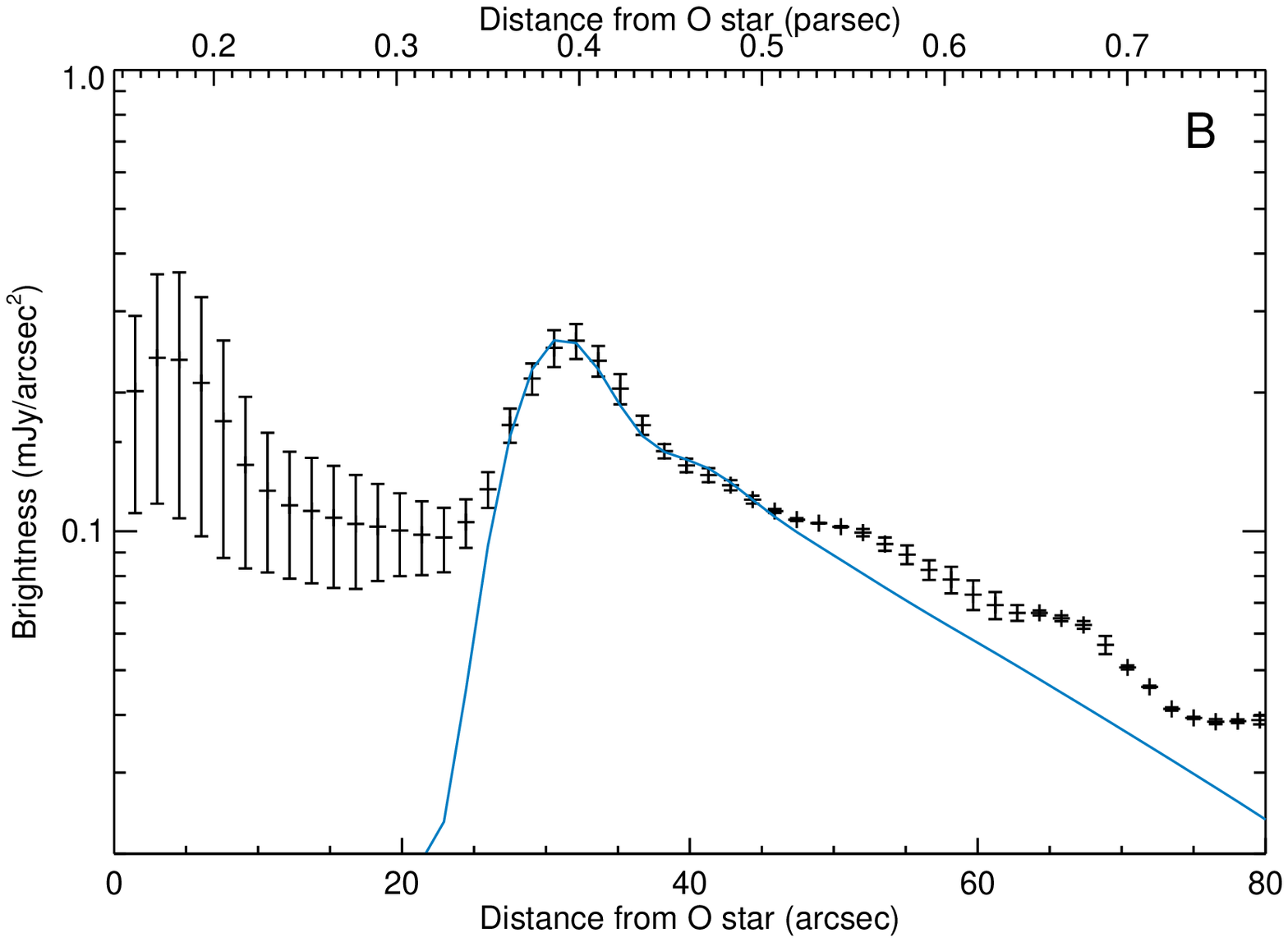}
\includegraphics*[width=3.2in]{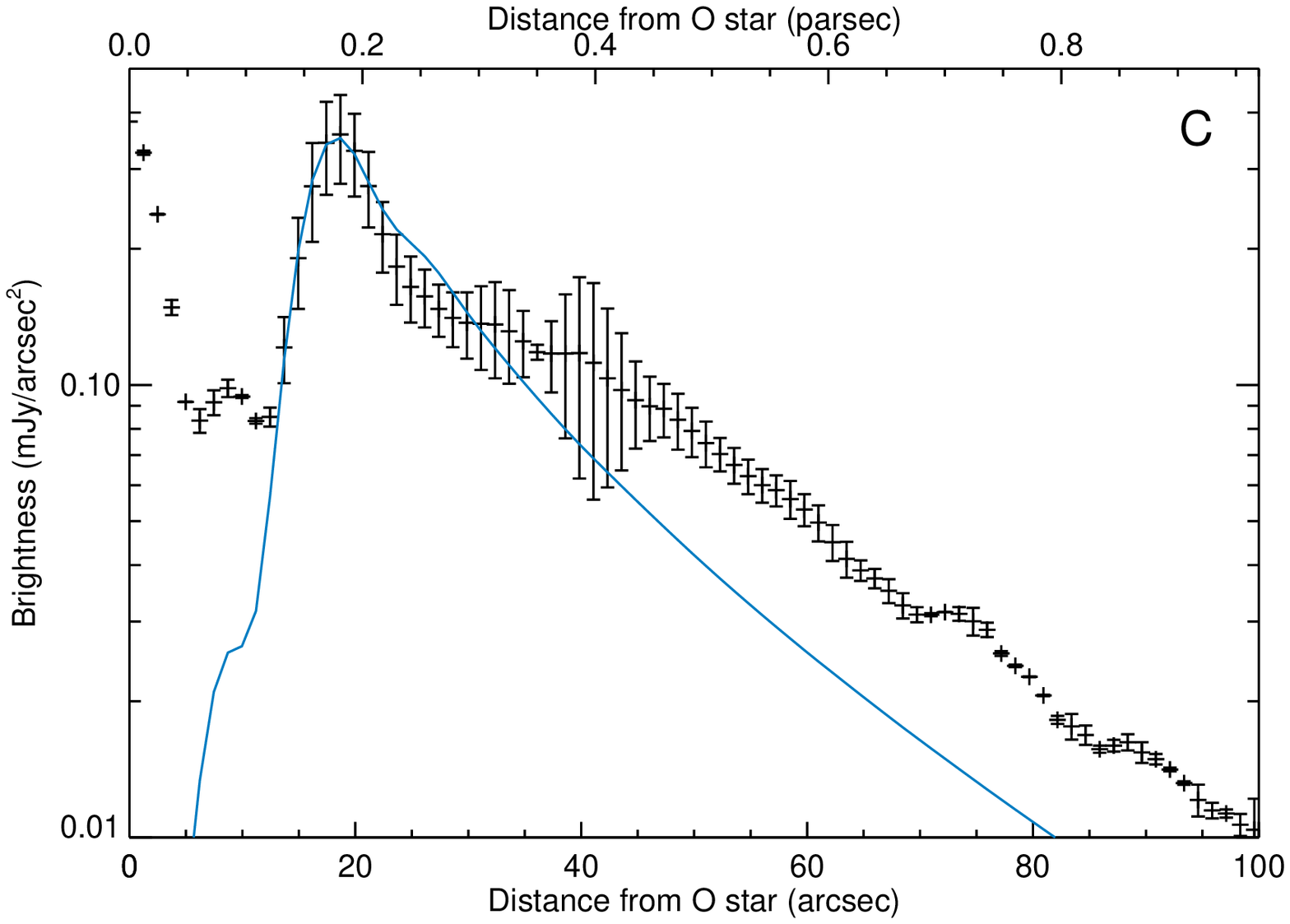}
\includegraphics*[width=3.2in]{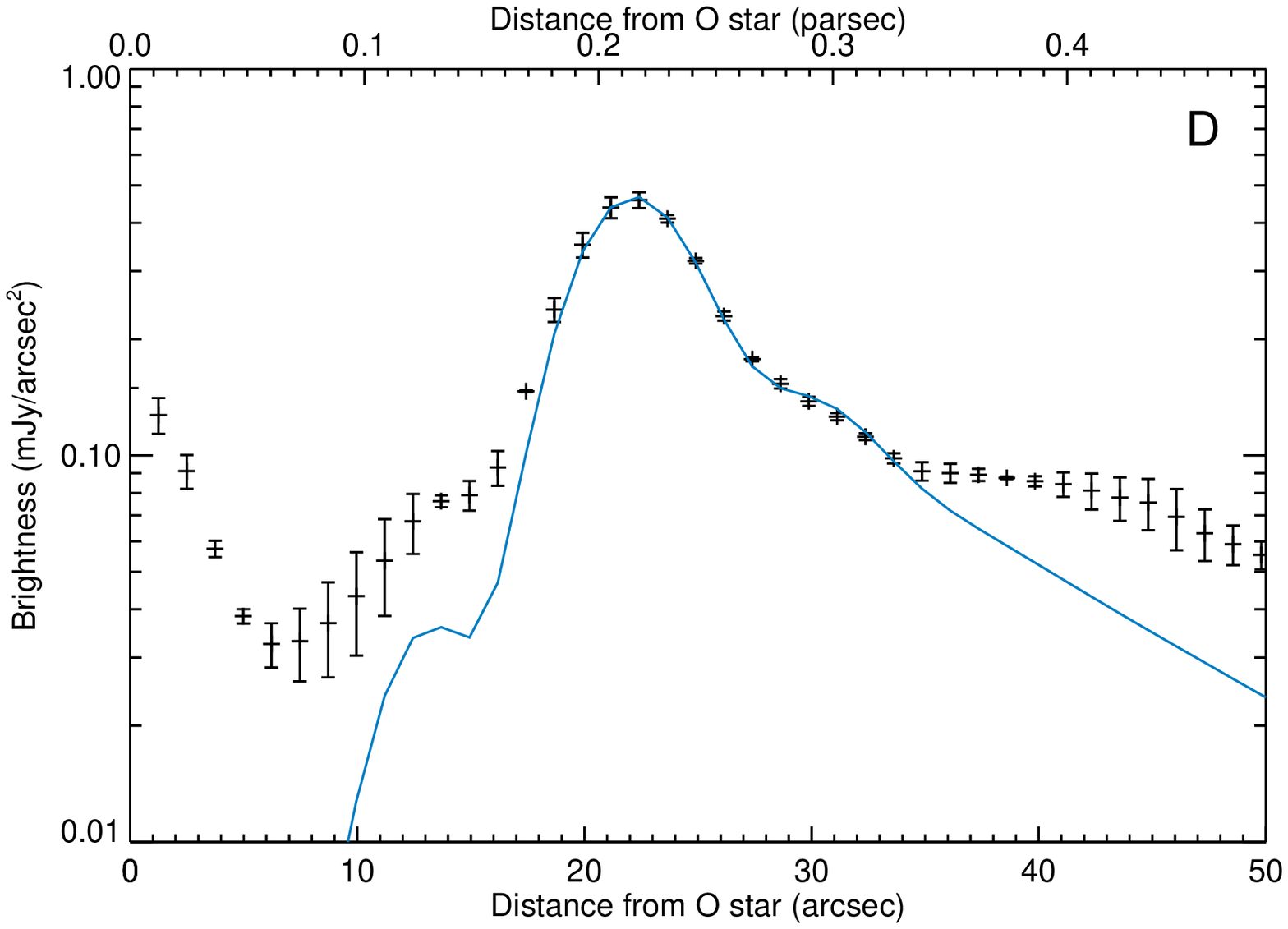}
\caption{Fits to the 24 $\micron$ brightness distributions of the four
  cometary tails in W5. Contamination from neighboring sources and
  surrounding diffuse emission likely causes much of the discrepancy
  between the data and the fits to tails B, C and D. Data points are
  shown by diamond points with error bars. The best fit is the solid
  blue line passing through the points.}
\end{center}
\end{figure}

\end{document}